\documentstyle[editedbook,epsfig,psfig,epsf]{mq} 

\def\cm2{cm$^2$ } 
\def\se1{s$^{-1}$ }

\begin{opening} 
\title{The atoll source 4U 1608--52 is not a Z source!} 
\author{S. van Straaten$^1$, M. van der Klis$^1$ and M. M\'endez$^2$} 
\institute{
$^1$ Astronomical Institute, ``Anton Pannekoek'',
University of Amsterdam, and Center for High Energy Astrophysics, 
Kruislaan 403, 1098 SJ Amsterdam, The Netherlands.\\ 
$^2$ SRON, National Institute for Space Research, Sorbonnelaan 2, 
3584 CA Utrecht, The Netherlands.} 
\end{opening} 
 
\runningtitle{The atoll source 4U 1608--52 is not a Z source!} 
\runningauthor{van Straaten, van der Klis \& M\'endez} 
 
\begin{document} 
\vspace{-0.5cm} 
\begin{abstract} 
{\small We have studied the spectral and timing behaviour of the atoll 
source 4U 
1608--52. We find that the timing behaviour of 4U 1608--52 is almost 
identical to that of the atoll sources 4U 0614+09 and 4U 1728--34. Recently 
Muno, Remillard \& Chakrabarty (2002) and Gierlinski \& Done (2002) suggested 
that the atoll sources trace out similar three--branch patterns as the Z 
sources. The timing behaviour is not consistent with the idea that 4U 1608--52 
traces out a three--branched Z shape in the color--color diagram along which 
the timing properties vary gradually.} 
\end{abstract} 
 
\section{Introduction and Data Analysis} 
In this work we use all available data from RXTE's PCA to simultaneously study 
the spectral and timing properties in the transient low mass X-ray binary 
4U 1608--52. We calculate a color--color diagram. As the energy spectrum of a source changes, it 
moves through the diagram. To study the timing we calculate Fourier power
density spectra and fit them with the 
multi--Lorentzian fit function; a sum of 
Lorentzian components plus an occasional power law to fit the very low 
frequency noise \cite{bpk02,vstr02}. It has been recently proposed 
\cite{muno02,gier02} that the atoll sources trace out similar three--branch 
patterns as the Z sources; one of our
goals in this work is to test this hypothesis.\\

\section{Stepping through the color--color diagram} 
As a first step we go through the data in chronological order and look at the 
timing properties (per observation) 
and the position of the source in the color--color diagram.
The obtained lightcurve and color--color 
diagram for 4U 1608--52 can be grouped into 3 parts. The first part ranges 
from 1996 March 3 to December 28 (the decay of the 1996 outburst, 
see \cite{berg96}), the second part from 1998 February 3 to September 29 
(the 1998 outburst, see \cite{mendez98}) and the third part from 2000 
March 6 to May 10. In practice most data was available for the second part 
of the lightcurve (the 1998 outburst) so we will present the results for the 
second part of the lightcurve first. The results can then serve as a template 
for the rest of the data.

For the 1998 outburst we find 7 different color diagram position/power 
spectral classes and we can confirm the result of \cite{mendez98} that the 
color--color diagram shows the classical atoll C shape (see \cite{hk89}). 
For the decay of the 1996 outburst we find one additional class that was 
different from those observed during the 1998 outburst. The color--color 
diagram deviates from the C shape and if we sort the 
classes by characteristic frequency, the color--color diagram 
seems to follow an $\epsilon$ shape instead of the classical atoll shape 
(see also below).  
In the third part of the data the source countrates were low and in most 
cases it was impossible to identify any power spectral features, therefore 
the classification for this part of the data was solely done on position in 
the color--color diagram.

\section{Combining the power spectra} 
 
\begin{figure}[htb] 
\centering \epsfig{file=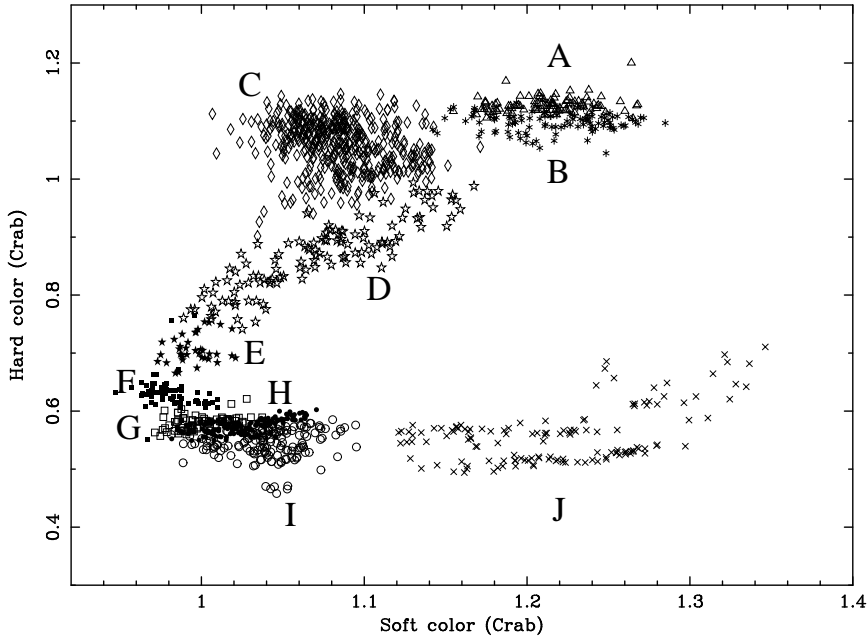,width=11.5cm}
\caption{Color--color diagram divided into 10 representative intervals 
marked from A to J.}
\label{fig:cc_int}
\end{figure}

To improve the statistics of the power spectra we add up all the continuous 
time intervals in each of the 8 classes. To avoid doubling of the lower 
kilohertz QPO peak we split the class marked with the 
filled circles up into three parts depending on lower kilohertz QPO 
frequency. In Figure \ref{fig:cc_int} we show the resulting 10 intervals 
marked from A to J in the color--color diagram.\\ 

\begin{figure}[htb] 
\centering \epsfig{file=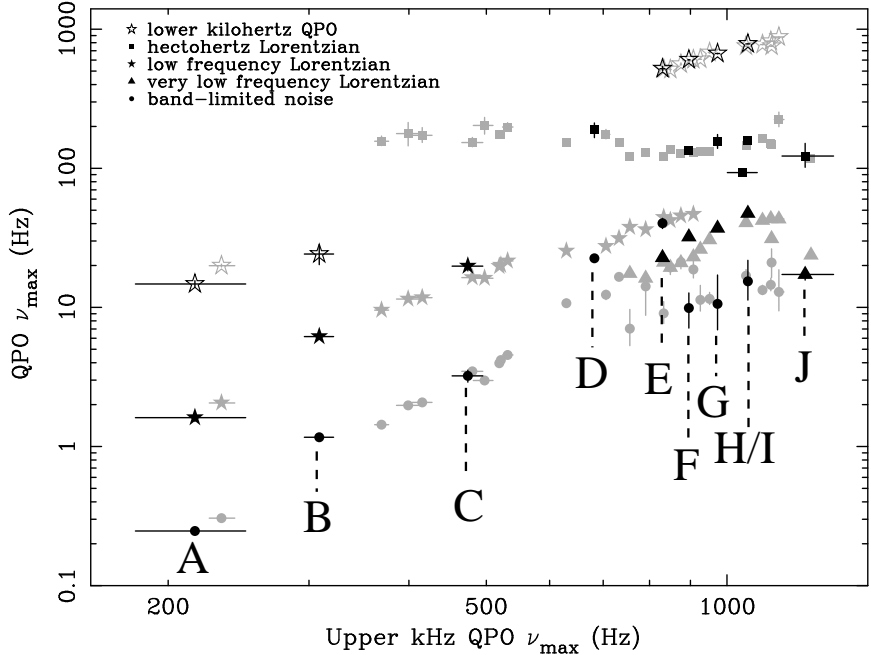,width=11.5cm}
\caption{Correlations between the characteristic frequencies 
 of the several Lorentzians used to 
fit the power spectra of 4U 1608--52 and the 
characteristic frequency of the Lorentzian identified as the upper 
kilohertz QPO. The black points mark the results 
for 4U 1608--52, the grey points the results for 4U 1728--34 and 4U 0614+09. 
The symbols mark the different power spectral components. The letters mark 
the intervals of Figure \ref{fig:cc_int}. Note that the identification of 
the lower kilohertz QPO at low frequencies (intervals A and B) is uncertain.}
\label{fig:freq_freq}
\end{figure}

We fit the power density spectrum of each interval of Figure \ref{fig:cc_int} 
with the multi--Lorentzian fit function. 
In Figure \ref{fig:freq_freq} we show the characteristic frequencies 
of the Lorentzians used to fit the power spectra of 4U 1608--52 
plotted versus the characteristic frequency of the Lorentzian identified as 
the upper kilohertz QPO, together with the results of \cite{vstr02} for 
4U 1728--34 and 4U 0614+09.  The results of 
the multi--Lorentzian fit to 4U 1608--52 are remarkably similar to those of 
4U 1728--34 and 4U 0614+09.

Interval C in Figure \ref{fig:cc_int} represents a deviation from the 
classical atoll shape (see also \cite{muno02} and \cite{gier02}). According to 
the interpretation of \cite{muno02} and \cite{gier02} interval C would 
represent the analogon to the horizontal branch of the Z sources. 
As in Z sources the characteristic frequencies of the timing features 
increase along the Z starting at the horizontal branch, the measured 
frequencies in C should then be lower than those in A and B. Instead we find 
that all characteristic frequencies clearly increase from A to J so that the
frequencies in C are intermediate between those in B and D. This 
is not consistent with the idea that 4U 1608-52 traces out a three-branched 
Z shape in the color-color diagram, 

\section*{Acknowledgements} 
This work was supported by NWO SPINOZA grant 08--0 to E.P.J. van den Heuvel, 
by the Netherlands Organization for Scientific Research (NWO), and by 
the Netherlands Research School for Astronomy (NOVA). 
This research has made use of data obtained through
the High Energy Astrophysics Science Archive Research Center Online Service, 
provided by the NASA/Goddard Space Flight Center.

\end{document}